\documentclass[11pt]{article}
\usepackage{mathptmx}
\usepackage{geometry}                
\usepackage{graphicx}
\usepackage{amssymb}
\usepackage{amsmath}
\usepackage{amsthm}
\usepackage{graphics}

\def\Q{\mathbb{Q}}
\def\R{\mathbb{R}}
\def\N{\mathbb{N}}

\def\Z{\mathbb{Z}}

\def\deq{\stackrel{\mathrm{def}}{=}}

\def\ZB{\Z_{\beta}}
\newtheorem{thm}{Theorem}[section]

 \newtheorem{corollary}[thm]{Corollary}
\newtheorem{remark}[thm]{Remark}
\newtheorem{proposition}[thm]{Proposition}
\newtheorem{example}[thm]{Example}
\title{Asymptotic behavior of beta-integers}

\author{L. Balkov\'a$^{\mathrm{a},\mathrm{b}}$,  J. P. Gazeau$^{\mathrm{b}}$,
E. Pelantov\'a$^{\mathrm{a}}$  \footnote{e-mail: l.balkova@centrum.cz,gazeau@apc.univ-paris7.fr, edita.pelantova@fjfi.cvut.cz}\\
\emph{$^{\mathrm{a}}$ Department of Mathematics, FNSPE,
Czech Technical University},\\ \emph{
Trojanova 13, 120 00 Praha 2, Czech Republic}\\
\emph{$^{\mathrm{b}}$ Laboratoire APC,
Universit\'e Paris 7-Denis Diderot,}   \\
\emph{10, rue A. Domon  et  L. Duquet
75205 Paris Cedex 13, France}}

\begin{document}

\maketitle

\begin{abstract}
Beta-integers (``$\beta$-integers'') are those  numbers which are
the counterparts of integers when real numbers are expressed in
irrational basis $\beta > 1$. In quasicrystalline studies
$\beta$-integers  supersede the ``crystallographic'' ordinary
integers. When the number $\beta$ is a   Parry number, the
corresponding $\beta$-integers realize only a finite number of
distances between consecutive elements and somewhat appear like
ordinary integers, mainly in an asymptotic sense. In this letter we
make  precise this asymptotic behavior by proving four theorems
concerning Parry $\beta$-integers.
\end{abstract}

\section{Introduction: $\beta$-integers versus integers in quasicrystals}

Aperiodicity of quasicrystals~\cite{jan} implies the absence of
space-group symmetry to which integers are inherent~\cite{IUC}. On
the other hand, experimentally observed quasicrystals show
self-similarity (\emph{e.g.} in their diffraction pattern). The
involved self-similarity factors  are quadratic Pisot-Vijayaraghavan
(PV) units. We recall that an algebraic integer $\beta>1$ is called
a PV number if all its Galois conjugates have modulus less than one.
Mostly observed factors are the \emph{golden mean}
$\tau=\tfrac{1+\sqrt5}{2}$ and its square
$\tau^2=\tfrac{3+\sqrt5}{2}$, associated to decagonal and
icosahedral quasicrystals respectively. The other ones are
$\delta=1+\sqrt2$ (octagonal symmetry) and $\theta={2+\sqrt3}$
(dodecagonal symmetry). Each of these numbers, here generically
denoted by  $\beta$, determines a~discrete set of real numbers,
$\Z_\beta = \{b_n \mid
 n \in \Z \}$ \cite{BuFrGaKr}.  The set of
$\beta$-integers $b_n$ aims to play the role that integers play in
crystallography. In the same spirit, $\beta$-lattices $\Gamma$ are
based on $\beta$-integers, like lattices are based on integers:
$\Gamma=\sum_{i=1}^d\ZB{\bf e}_i$,  with $({\bf e}_i)$ a base of
$\R^d$. These point sets in $\R^d$ are eligible frames in which one
could think of the properties of quasiperiodic point-sets and
tilings, thus generalizing the notion of lattice in periodic cases
\cite{EFGV}. As a matter of fact, it has become like a paradigm that
geometrical supports of quasi-crystalline structures should be
Delaunay sets obtained through ``Cut-and-Projection'' from
higher-dimensional lattices. Now, it can be proved \cite{FGK} that
most of such Cut and Project sets are subsets of suitably rescaled
$\beta$-lattices.

$\beta$-integers and their $\beta$-lattice extensions have
interesting arithmetic and diffractive properties which generalize
to some extent the additive, multiplicative and diffractive
properties of integers \cite{EFGV,GoGa, GaGo}. Throughout these
features, a question  arises to which the content of this paper will
partially answer. In view of the ``quasi''-periodic distribution of
the $b_n$'s on the real line, it is natural to investigate the
extent to which they differ from ordinary integers, according to the
nature of $\beta$. If $\beta$ is endowed with specific algebraic
properties, one will see that the corresponding set of $\beta$-integers is a mathematically
``well-controlled'' perturbation of $\Z$ in an asymptotic sense
\begin{equation}
\label{asbetaint} b_n \underset{n \to \infty}{\approx} c_{\beta}\,
(n + \alpha(n))\,
\end{equation}
where $n \mapsto \alpha(n)$ is a~bounded sequence and the scaling
coefficient $c_{\beta}$ is in $\Q(\beta)$. As a matter of fact, for a large
family of numbers $\beta$, namely the \emph{Parry} family,
$\beta$-lattices asymptotically appear as lattices. Parry numbers
give rise to $\beta$-integers which realize only a~finite number of
distances between consecutive elements and so appear as the most comparable  to ordinary integers.

We give in the present letter a set of rigorous results concerning
the asymptotic behavior of $\beta$-integers  when $\beta$ is
a~generic Parry number. More precisely, we will illustrate the
similarity between sets $\mathbb N$ and $\mathbb Z^{+}_\beta=\{b_n
\mid n \in \mathbb N\}$ for $\beta$ being a~Parry number by proving
two properties:
\begin{enumerate}
\item We will show that $c_{\beta}=\lim_{n \to \infty}\dfrac{b_n}{n}$ exists
and we will provide a~simple formula for $c_{\beta}$.
\item For $\beta$ being moreover a~PV number with mutually
distinct roots of its Parry polynomial, we will prove that
$(b_n-c_{\beta}\,n)_{n \in \mathbb N}$ is a~bounded sequence.
\end{enumerate}

Let us mention that both of the previous asymptotic characteristics
are known for $\beta$ being a~quadratic unit. The following
proposition providing explicit formulae for $\beta$-integers for
a~quadratic unit $\beta$ comes from~\cite{GaGo}.
\begin{proposition}\label{unit_gago} If $\beta$ is a~simple
quadratic Parry unit, then
$$
 {\mathbb Z}_{\beta}^+ =  \left\{ b_n = c_{\beta} n  +
\frac{1}{\beta} \, \frac{1 - \beta}{1 + \beta} +
\frac{\beta-1}{\beta}\, \left\{ \frac{n+1}{ 1 + \beta}\right\}, \  n
\in \mathbb N \right\}, \quad \text{where} \ c_{\beta}= \dfrac{1 +
\beta^2}{\beta (1 + \beta)}.
$$

\vspace{0.1cm} If $\beta$ is a~non-simple quadratic Parry unit, then
$$
{\mathbb Z}_{\beta}^+ =   \left\{ b_n = c_{\beta} n  +
\frac{1}{\beta} \left\{ \frac{n}{  \beta}\right\}, \  n \in \mathbb
N \right\}, \quad \text{where} \ c_{\beta} = 1 - \dfrac{1}{\beta^2}\, ,
$$
where $\{ x\}$ designates the fractional part of a nonnegative real number $x$.
\end{proposition}

 In Section 2, we provide a~brief but
self-contained description of the concept of $\beta$-integers, with
emphasis on $\beta$-integers realizing only a~finite number of
distances between neighbors, i.e., $\beta$-integers associated with
Parry numbers $\beta$. The distances are then coded by letters, and,
thanks to the self-similarity of $\mathbb Z_\beta$, the obtained
infinite words $u_\beta$ are known to be fixed points of
substitutions. As the associated substitution matrices are
primitive, the Perron-Frobenius theorem together with some further
applications from matrix theory will lead to the  results in Section 3 on the asymptotic behavior of $\beta$-integers for Parry numbers $\beta$.

\section{All we need to know on $\beta$-integers}\label{beta}

\subsection{$\beta$-representation and $\beta$-expansion}\label{beta_repr}
Let $\beta >1$ be a~real number and let $x$ be a~non-negative real number. Any
convergent series of the form $ x=\sum_{i=-\infty}^{k} x_i\beta^{i}
\stackrel{\mathrm{not}}{=}x_k x_{k-1}\cdots x_0 \bullet
x_{-1}\cdots$, where $x_i \in \mathbb N \stackrel{\mathrm{def}}{=}
\{0,1,2,\dotsc\}$ and $x_k \not =0$, is called a~{\em
$\beta$-representation} of $x$. Let $\beta$ be a~positive integer,
then if we admit only $\{0,1,\dots, \beta-1\}$ as the set of
coefficients and if we avoid the suffix $(\beta-1)^{\omega}$, where
$\omega$ signifies an infinite repetition, there exists a~unique
$\beta$-representation for every $x$, called the {\em standard}
$\beta$-representation. For $\beta=10$ (resp. $\beta=2$), it is the
usual {\em decimal} (resp. {\em binary}) representation.

Even if $\beta$ is not an integer, every positive
number $x$ has at least one $\beta$-representation. This
representation can be obtained by the following {\it greedy
algorithm:}
\begin{enumerate}
\item Find  $k \in \mathbb Z$ such that $\beta^{k} \leq x <
\beta^{k+1}$ and put $x_k:=\left\lfloor
\dfrac{x}{\beta^{k}}\right\rfloor$ and $r_k:=\left\{
\dfrac{x}{\beta^{k}}\right\}$, where $\left\lfloor x
\right\rfloor=x-\{x\}$ denotes the lower integer part of $x$.
\item For $i<k$, put $x_i:=\lfloor \beta r_{i+1}\rfloor$ and
$r_i:=\{\beta r_{i+1}\}$.
\end{enumerate}

This representation  is called the {\em $\beta$-expansion} of $x$
and the coefficients of the~$\beta$-expansion clearly satisfy:
$x_{k} \in \{1,\ldots,\lfloor \beta \rfloor\}$ and $x_i \in
\{0,\ldots,\lfloor \beta \rfloor\}$ for all $i<k$. We use the
notation ${\langle x \rangle} _{\beta}$ for the $\beta$-expansion of
$x$. For $\beta$ being an integer, the $\beta$-expansion coincides
with the standard $\beta$-representation.

The greedy algorithm implies that among $\beta$-representations, the
$\beta$-expansion is the largest according to the radix order.
Actually the latter corresponds to the
ordering of real numbers: for all $x,y \in [0,+\infty)$, the
inequality $x<y$ holds if and only if ${\langle x \rangle}_\beta$ is
smaller than ${\langle y \rangle}_\beta$ according to the radix
order.
\begin{example}
Let $\beta=\tau=\frac{1+\sqrt{5}}{2}$.  The golden mean $\tau$ is
the larger root of the polynomial $x^2-x-1$. Applying the greedy
algorithm, we get for instance the following $\tau$-expansions:\\
${\langle \frac{\sqrt{5}-1}{2}\rangle}_{\tau}=0\bullet 1, \ {\langle
\frac{3+\sqrt{5}}{2} \rangle}_{\tau}=100\bullet, \ {\langle
\frac{5+3\sqrt{5}}{10} \rangle}_{\tau}=1\bullet
(0001)^{\omega}\cdots$.
\end{example}

\subsection{R\'enyi expansion of unity}
The $\beta$-expansion of numbers from the interval $[0,1)$ can be
obtained through the map  $T_\beta:
[0,1]\rightarrow [0,1)$ defined by
\begin{equation}\label{T_beta}
T_{\beta}(x)=\{\beta x\}.
\end{equation} It is easy to verify that
for every $x \in [0,1)$, it holds ${\langle x
\rangle}_{\beta}=0\bullet x_{-1} x_{-2}\cdots$ if and only if
$x_{-i}=\lfloor \beta T_{\beta}^{-(i+1)}(x) \rfloor.$ By extending
Formula~\eqref{T_beta} to $x=1$, one gets the so-called R\' enyi
expansion of unity ~\cite{Re} in base $\beta$:
\begin{equation}\label{d_beta_1}
d_{\beta}(1)=t_1t_2t_3\cdots, \quad \mbox{where }\quad t_i:=\lfloor
\beta T_{\beta}^{i-1}(1)\rfloor. \end{equation}
Every number
$\beta>1$ is characterized by its R\'enyi expansion of unity. Note
that $t_1=\lfloor \beta\rfloor \geq1$. Parry in~\cite{Pa} has moreover shown that the
R\'enyi expansion of unity enables us to decide whether a~given
$\beta$-representation of $x$ is its $\beta$-expansion or not. For
this purpose, we define the {\em infinite R\' enyi expansion of
unity} (it is the largest infinite $\beta$-representation of 1 with
respect to the radix order)
\begin{equation}\label{infinite_exp}
d^{*}_{\beta}(1)= \left\{\begin{array}{ll} d_{\beta}(1) & \hbox{if}
\quad d_{\beta}(1) \quad \hbox{is infinite},\\
(t_1 t_2\cdots t_{m-1}(t_m -1))^{\omega}
 & \hbox{if}
\quad d_{\beta}(1)=t_1\cdots t_m \quad \hbox{with} \quad t_m \not =
0.
\end{array}\right.
\end{equation}
\begin{proposition}[Parry condition] \label{Parry_expansions}
Let $d^{*}_{\beta}(1)$ be the infinite R\'enyi expansion of unity in
base $\beta$. Let $\sum_{i=-\infty}^{k} x_i\beta^{i}$ be
a~$\beta$-representation of a~non-negative number $x$. Then
$\sum_{i=-\infty}^{k} x_i\beta^{i}$ is the $\beta$-expansion of $x$
if and only if
\begin{equation}\label{ParryCondition}
x_ix_{i-1}\cdots \prec d^{*}_{\beta}(1) \ \mbox{ for all }\ i\leq k,
\end{equation}
where $\prec$ means smaller in the lexicographical order.
\end{proposition}
\begin{example}
For $\beta=\tau=\frac{1+\sqrt{5}}{2}$, the R\'enyi expansion of
unity is $d_\tau(1)=11$. Then, $d^{*}_\tau(1)=(10)^{\omega}$, and,
according to the Parry condition, any sequence of coefficients in
$\{0,1\}$ which does not end with $(10)^{\omega}$ and which does not
contain the block 11 is the $\tau$-expansion of a~non-negative real
number.
\end{example}

\subsection{Parry numbers}\label{algebraic_numbers}
This paper is concerned by  real numbers $\beta>1$ having an
eventually periodic R\'enyi expansion of unity $d_\beta(1)$, i.e.,
$d_\beta(1)=t_1\dots t_m(t_{m+1}\dots t_{m+p})^{\omega}$ for some
$m,p \in \mathbb N$, called {\em Parry numbers}. For every Parry
number $\beta$, it is easy to recover, from the eventual periodicity
of $d_\beta(1)$, a~monic polynomial with integer coefficients having
$\beta$ as a~root, i.e., $\beta$ is an algebraic integer. However,
this so-called {\em Parry polynomial} is not necessarily the minimal
polynomial of $\beta$. It was proven in~\cite{Bertrand} that every
Pisot number is a~Parry number.
\subsection{Definition and properties of $\beta$-integers}
 Nonnegative numbers $x$ with vanishing
$\beta$-fractional part are called {\em nonnegative
$\beta$-integers}, formally,
$
{\mathbb Z}_{\beta}^{+}:=\{x \geq 0 \bigm | {\langle x \rangle}
_{\beta}=x_k x_{k-1}\cdots x_0 \bullet \}.
$
The set of {\em $\beta$-integers} is then defined by
$
{\mathbb Z}_{\beta}:=\bigl( -{\mathbb Z}_{\beta}^{+}\bigr) \cup
{\mathbb Z}_{\beta}^{+}.
$
As already mentioned, the radix order on $\beta$-expansions
corresponds to the natural order of non-negative real numbers.
Consequently, there exists a~strictly increasing sequence $(b_n)_{n
=0}^{\infty}$ such that
\begin{equation}\label{b_n}
b_0=0 \quad \mbox{and} \quad \{b_n \bigm | n \in \mathbb
\N\}=\mathbb Z_\beta^{+}.
\end{equation}
When  $\beta$ is an integer, $\mathbb Z_\beta=\mathbb Z$ and so the
distance between the neighboring elements of $\mathbb Z_{\beta}$ is
always~1. The situation changes dramatically if $\beta \not \in
\mathbb N$. In this case, the number of different distances between
the neighboring elements of $\mathbb Z_{\beta}$ is at least 2 and
the set $\mathbb Z_{\beta}$ keeps only partially the resemblance to
$\mathbb Z$:
\begin{enumerate}
\item
$\mathbb Z_{\beta}$ has no accumulation points.
\item
$\mathbb Z_{\beta}$ is relatively dense, i.e., the distances between
consecutive elements of $\mathbb Z_{\beta}$ are bounded.
\item
$\mathbb Z_{\beta}$ is self-similar, thus $\beta \mathbb
Z_{\beta}\subset \mathbb Z_{\beta}$.
\item
$\mathbb Z_{\beta}$ is not invariant under translation.
\item
$\mathbb Z_\beta$ forms a~Meyer set if $\beta$ is a~Pisot number,
i.e., $\mathbb Z_{\beta}-\mathbb Z_{\beta} \subset \mathbb
Z_{\beta}+F$ for a~finite set $F\subset \mathbb R$ (proved
in~\cite{BuFrGaKr}).
\end{enumerate}
Thurston~\cite{Th} has shown that distances occurring between
neighbors of $\mathbb Z_{\beta}$ form the set
$\mbox{$\{\Delta_{k}\bigm | k \in \mathbb N \}$}$, where
\begin{equation}\label{distance_beta}
\Delta_{k}:=\sum_{i=1}^{\infty}\frac{t_{i+k}}{{\beta}^{i}} \ \ \
\hbox{for} \ k \in \mathbb N\,.
\end{equation}
It is evident that the set $\{\Delta_{k}\bigm | k \in \mathbb N
\}$ is finite if and only if $d_{\beta}(1)$ is eventually periodic.

\subsection{Infinite words and substitutions associated with $\beta$-integers}
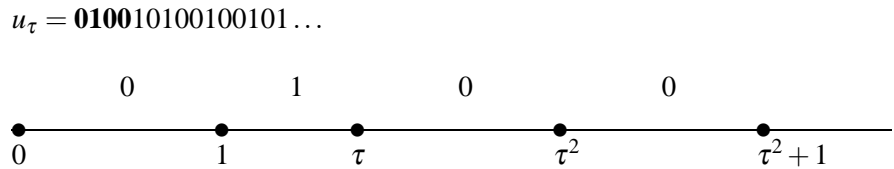
\begin{figure}[ht]
\vskip 0.5cm
\begin{center}
\unitlength=0.9cm
\begin{picture}(8.5,1)

\put(-1.6,1){$u_\tau={\bf 0100}10100100101\dots$}
\put(-1.6,-0.5){\line(1,0){13}}

\put(-1.5,-0.5){\circle*{.2}} \put(1.5,-0.5){\circle*{.2}}
\put(3.5,-0.5){\circle*{.2}} \put(6.5,-0.5){\circle*{.2}}
\put(9.5,-0.5){\circle*{.2}}

 \put(0,0){$0$}
\put(2.5,0){$1$} \put(5,0){$0$}\put(8,0){$0$}

\put(-1.6,-1){$0$} \put(1.4,-1){$1$} \put(3.4,-1){$\tau$}
\put(6.4,-1){$\tau^2$}\put(9.4,-1){$\tau^2+1$}
\end{picture}
\end{center}
\vskip 0.5cm \caption{First elements of $\Z_\tau^+$ (non-negative
$\tau$-integers) and of the associated infinite word
$u_\tau$.}\label{tauinteg}
\end{figure}

If the number of distances between neighbors in $\mathbb Z_\beta$ is
finite, one can associate with every distance a~letter. Thus, we
obtain an infinite word $u_\beta$ coding $\mathbb Z_\beta^{+}$ as
illustrated in Figure~\ref{tauinteg}.
In order to define $u_\beta$ properly, let us introduce some
definitions from combinatorics on words.
An {\em alphabet} $\mathcal A$ is a~finite set of symbols, called
{\em letters}. A~concatenation of letters is a~{\em word}. The set
$\mathcal A^{*}$ of all finite words (including the empty word
$\varepsilon$) provided with the operation of concatenation is
a~free monoid. The length of a~word $w=w_0w_1w_2\cdots w_{n-1}$ is
denoted by $|w|=n$ and $|w|_a$ denotes the number of letters $a\in
\mathcal A$ in $w$. We will deal also with infinite words
$u=u_0u_1u_2\cdots \in {\mathcal A}^{\mathbb N}$. A~finite word $w$
is called a~{\em factor} of the word $u$ (finite or infinite) if
there exist a~finite word $w^{(1)}$ and a~word $w^{(2)}$ (finite or
infinite) such that $u=w^{(1)}ww^{(2)}$. The word $w$ is a~{\em
prefix} of $u$ if $w^{(1)}=\varepsilon$.
A~{\em substitution} over $\cal A^{*}$ is a~morphism $\varphi:{\cal
A^{*}} \rightarrow {\cal A^{*}}$ such that there exists a~letter $a
\in \cal A$ and a~non-empty word $w \in {\cal A}^{*}$ satisfying
$\varphi(a)=aw$ and $\varphi$ is {\em non-erasing}, i.e.,
$\varphi(b) \not = \varepsilon$ for all $b \in \cal A$. Since any
morphism satisfies $\varphi(vw)=\varphi(v)\varphi(w)$ for all $v,w
\in {\cal A^{*}}$, the substitution is uniquely determined by images
of letters. Instead of classical $\varphi(a)=w$, we sometimes write
$a \to w$. A~substitution can be naturally extended to infinite
words $u \in {\mathcal A}^{\mathbb N}$ by the prescription
$\varphi(u)=\varphi(u_0)\varphi(u_1)\varphi(u_2)\dots$ An infinite
word $u$ is said to be a~{\em fixed point} of the substitution
$\varphi$ if it fulfills $u=\varphi(u)$. It is obvious that
a~substitution $\varphi$ has at least one fixed point, namely
$\lim_{n \to \infty}\varphi^{n}(a)$.
Morphisms on ${\mathcal A}^{*}$ equipped with the operation of
composition form a~monoid. Substitutions do not form a~monoid as the
identity is not a~substitution. Nevertheless, if $\varphi$ is
a~substitution, $\varphi^n$ is also a~substitution.

The following prescription associates with every substitution
$\varphi$ defined over the alphabet ${\mathcal A}=\{a_1, a_2, \dots,
a_d\}$ a~non-negative integer $d \times d$ matrix, called the {\em
substitution matrix} $M_{\varphi}$
\begin{equation}\label{matrix_subst}
\left(M_{\varphi}\right)_{ij}= |\varphi(a_i)|_{a_j}, \quad i,j \in
\{1, \dots, d\}\,.
\end{equation}
As an immediate consequence of the definition, it holds for any word
$w$ that
\begin{equation}\label{subst_length}
(|w|_{a_1}, |w|_{a_2}, \dots, |w|_{a_d})
M_\varphi=(|\varphi(w)|_{a_1}, |\varphi(w)|_{a_2}, \dots,
|\varphi(w)|_{a_d}).
\end{equation}
The substitution matrix of the composition of substitutions
$\varphi, \ \psi$ obeys the formula $M_{\varphi \circ
\psi}=M_{\psi}M_{\varphi}$.
\begin{example}\label{Fib_matrix} The Fibonacci substitution $\varphi$ defined on
$\{0,1\}$ by $\varphi(0)=01, \ \varphi(1)=0$, has the following
substitution matrix
\[M_{\varphi}=\left(\begin{array}{cc}
1 & 1  \\
1 & 0
\end{array}\right).\]
\end{example}

Let us finally recall
a~notion from matrix theory which proves useful in the study of
substitutions. A~matrix $M$ is called {\em primitive} if there
exists $k \in \mathbb N$ such that all entries of $M^k$ are
positive. There exists a~powerful theorem treating primitive matrices.
\begin{thm}[Perron-Frobenius]
Let $M$ be a~$d \times d$ primitive matrix. Then:
\begin{enumerate}
\item~\label{lambda} the matrix $M$ has a~positive eigenvalue $\lambda$ which is
strictly greater than the modulus of any other eigenvalue,
\item the eigenvalue $\lambda$ is algebraically simple,
\item to this eigenvalue corresponds a positive eigenvector (i.e., with positive entries only), while no other eigenvalue has a~positive eigenvector.
\end{enumerate}
\end{thm}
The eigenvalue $\lambda$ from the above theorem is called the {\em
Perron-Frobenius eigenvalue} of $M$. If a~primitive matrix $M$ is
the substitution matrix of a~substitution $\varphi$ over ${\mathcal
A}=\{a_1,\dots, a_d\}$, according to the result of
Queff\'elec~\cite{Qu}, the left eigenvector $(\rho_1, \ \rho_2,
\dots, \rho_d)$ of $\lambda$ normalized by $\sum_{i=1}^d \rho_i=1$
is equal to the vector of letter frequencies in any fixed point $u$
of $\varphi$, i.e.,
\begin{equation}\label{letter_frequencies} (\rho_1, \
\rho_2, \dots, \rho_d)=\bigl(\rho(a_1), \ \rho(a_2), \dots,
\rho(a_d)\bigr),
\end{equation}
where $\rho(a_i)=\lim_{n \to \infty}\dfrac{|u^{(n)}|_{a_i}}{n}$ and
$u^{(n)}$ denotes the prefix of $u$ of length~$n$.

\subsection{Parry numbers and the infinite words
$u_\beta$}\label{u_beta_non_quadratic}

From the formula for
distances~(\ref{distance_beta}), we know that the number of
distances in $\mathbb Z_\beta$ is finite if and only if the R\'enyi
expansion of unity $d_\beta(1)$ is eventually periodic, i.e., if
$\beta$ is a~Parry number.
\begin{itemize}
\item If $d_{\beta}(1)$ is finite, i.e., $d_{\beta}(1)=t_1t_2\dots t_m$, $t_m\not =0$, $\beta$ is said to be
a~{\em simple Parry number}, and the set of distances is
$\{\Delta_0, \Delta_1, \dots, \Delta_{m-1}\}$, where all of the
listed elements are mutually distinct.
\item If $d_\beta(1)$ is eventually periodic, but not finite, $\beta$ is a~{\em non-simple Parry number}.
Choose $p,m \in \mathbb N$ to be minimal such that
$d_\beta(1)=t_1t_2\dots t_m(t_{m+1}\dots t_{m+p})^{\omega}$, then
the set of all mutually distinct distances is $\{\Delta_0, \Delta_1,
\dots, \Delta_{m+p-1}\}$.
\end{itemize}

Let us precisely define the infinite word $u_\beta=u_0 u_1u_2 \dots$
associated with $\mathbb Z_\beta^+$ for a~Parry number $\beta$. Let
$\{\Delta_0, \dots, \Delta_{d-1}\}$ be the set of distances between
neighboring $\beta$-integers and let $(b_n)_{n=0}^{\infty}$ be as
defined in~\eqref{b_n}, then
\begin{equation}\label{u_beta}
u_n\deq i \quad \mbox{if} \quad b_{n+1}-b_n=\Delta_i.
\end{equation}
\subsection{Canonical substitutions for Parry numbers}
Fabre in~\cite{Fabre} has associated with Parry numbers canonical
substitutions in the following way.

Let $\beta$ be a~simple Parry number, i.e., $d_\beta(1) =t_1t_2\dots
t_m$, for $m\in \mathbb N$. Then the corresponding canonical
substitution $\varphi$ is defined over the alphabet $\{0, 1, \dots,
m-1\}$ by
\begin{equation}\label{subst1general}
\varphi(0)=0^{t_1}1, \,
\varphi(1)=0^{t_2}2, \,
\dotsc \, ,\varphi(m-2)=0^{t_{m-1}}(m-1),\,
\varphi(m-1)=0^{t_m}.
\end{equation}
Similarly, let $\beta$ be a~non-simple Parry number, i.e.,
$d_\beta(1)=t_1t_2\dots t_m(t_{m+1}\dots t_{m+p})^{\omega}$. The
associated canonical substitution $\varphi$ is defined over the
alphabet $\{0, 1, \dots, m+p-1\}$ by
\begin{align}\label{subst2general}
\nonumber \varphi(0) = &\,  0^{t_1}1, \,
\varphi(1)=0^{t_2}2, \,
\dotsc \, ,
\varphi(m-1)= 0^{t_m}m, \,
\dotsc \\
& \dotsc  \, ,\varphi(m+p-2)=0^{t_{m+p-1}}(m+p-1),\,
\varphi(m+p-1)=0^{t_{m+p}}m.
\end{align}
Each of these substitutions has a~unique fixed point $\lim_{n \to
\infty}\varphi^{n}(0)$. Moreover, this fixed point turns out to be
equal to $u_\beta$. It is readily seen that in both cases, the
substitution matrix $M_\varphi$ is primitive. The characteristic
polynomial of the substitution matrix $M_\varphi$ coincides with the
Parry polynomial of the number $\beta$. Hence, the Perron-Frobenius
theorem implies that the Parry number $\beta$ is a~simple root of
its Parry polynomial $p(x)$ and therefore $p'(\beta) \neq 0$.

\section{Asymptotic behavior of $\beta$-integers for Parry numbers}

In order to derive some information about asymptotic properties of
$\beta$-integers, let us recall the essential relation given
in~\cite{Fabre} between a~$\beta$-integer $b_n$ and its coding by
a~prefix of the associated infinite word $u_\beta$.
\begin{proposition}\label{b_n_varphi}
Let $u_\beta$ be the infinite word associated with a~Parry number
$\beta$ and let $\varphi$ be the associated substitution of $\beta$,
then for every $b_n \in {\mathbb Z_\beta}^+$ it holds that $\langle
b_n \rangle_{\beta}=a_{k-1}\dots a_1a_0\bullet \ $ if and only if $\
\varphi^{k-1}(0^{a_{k-1}}) \dots \varphi(0^{a_{1}})0^{a_0}$ is
a~prefix of $u_\beta$ of length $n$.
\end{proposition}

Since every prefix of $u_\beta$ codes a~$\beta$-integer $b_n$,
Proposition~\ref{b_n_varphi} provides us with the following
corollary.
\begin{corollary}
Let $w$ be a~prefix of $u_\beta$, then there exist $k \in \mathbb N$
and $a_0,\ a_1, \dots, a_{k-1} \in \mathbb N$ such that
$$w=\varphi^{k-1}(0^{a_{k-1}})\dots \varphi(0^{a_{1}})0^{a_0},$$
where $a_{k-1}\dots a_1a_0\bullet$ is the $\beta$-expansion of
a~$\beta$-integer.
\end{corollary}

Let us denote $U_i:=|\varphi^{i}(0)|$, then $(U_i)_{i =0}^{\infty}$
is a~canonical numeration system associated with the Parry number
$\beta$ (defined and called $\beta$- {\em numeration system}
in~\cite{Bertrand} and studied in~\cite{Fabre}). For details on
numeration systems consult~\cite{Lo1}. Proposition~\ref{b_n_varphi}
implies that the greedy representation of an integer $n$ in this
system is given by
\[n=\sum_{i=0}^{k-1}a_i U_i \quad \text{if} \quad b_n=\sum_{i=0}^{k-1}a_i \beta^{i}. \]
Applying~\eqref{matrix_subst}, the sequence $(U_i)_{i=0}^{\infty}$
may be expressed employing the substitution matrix $M$ of $\varphi$
in the following way
\begin{equation}\label{UiFormula}
U_i=\left( 1, 0, \dots, 0\right) M^{\ i}\left( 1, 1, \dots,
1\right)^T,
\end{equation}
where $T$ is for matrix transposition.
Let $\beta$ be a~simple Parry number, then the R\'enyi expansion of
unity is of the form $d_\beta(1)=t_1t_2\dots t_m$ and $\beta$ is the
largest root of the Parry polynomial
$p(x)=x^m-(t_1x^{m-1}+t_2x^{m-2}+\dots + t_{m-1}x+t_m)$. Let us
recall that $p(x)$ may be reducible.

\vspace{0.3cm} Our first aim is to provide a~simple formula for
constant $c_\beta$ such that $b_n \underset{n \to \infty}{\approx}
c_\beta n$. For any root $\gamma$ of $p(x)$, it is easy to verify
that $(\gamma^{m-1}, \ \gamma^{m-2}, \dots, \gamma, \ 1)$ is a~left
eigenvector of the substitution matrix $M$ associated with $\gamma$.
On the other hand, according to (\ref{letter_frequencies}), the
unique left eigenvector $(\rho_0, \rho_1, \dots, \rho_{m-1})$ of $M$
with $\sum_{i=0}^{m-1}\rho_i =1$ is such  that $\rho_i$ is the
frequency of letter $i$ in $u_\beta$. Combining the two previous
facts, we obtain for frequencies the following formula
\begin{equation}\label{frequency1}
\rho_i=\frac{\beta^{m-1-i}}{\sum_{i=0}^{m-1}\beta^{i}}.
\end{equation}

Let $(\Delta_0, \ \Delta_1, \dots, \Delta_{m-1})^T$ be the right
eigenvector of $M$ associated with $\beta$ such that $\Delta_0=1$,
then it is easy to verify that $\Delta_i$ is the distance between
consecutive $\beta$-integers which is coded by letter $i$ in the
infinite word $u_\beta$ (see the formula~(\ref{distance_beta}) for
distances). For our purposes, the following easily derivable formula
for distances will be useful
\begin{equation}\label{distance}
\Delta_i=\beta^{i}-\sum_{j=1}^{i}t_j \beta^{i-j}, \quad i \in \{0,1,
\dots, m-1\}.
\end{equation}

    \begin{thm}\label{c_formula}
    Let $p(x)$ be the Parry polynomial of a~simple Parry number $\beta$. Then
    $$c_{\beta}:=\lim_{n \to
    \infty}\frac{b_n}{n}=\frac{\beta-1}{\beta^m-1}p'(\beta).$$
    \end{thm}
    \begin{proof}
    Let us denote by $u$ the prefix of $u_\beta$ of length $n$, then
    $$b_n=|u|_0\Delta_0+|u|_1\Delta_1+\dots+ |u|_{m-1}\Delta_{m-1}.$$
    Since frequencies of letters exist, $\lim_{n \to
    \infty}\tfrac{b_n}{n}$ exists and obeys the following formula
    $$\lim_{n \to \infty}\frac{b_n}{n}=\rho_0\Delta_0+\rho_1\Delta_1+\dots+\rho_{m-1}\Delta_{m-1}.$$
    Applying~(\ref{frequency1}) and~(\ref{distance}), we obtain
    $$\begin{array}{rclc}\lim_{n \to \infty}\dfrac{b_n}{n}&=&\frac{1}{\sum_{i=0}^{m-1}\beta^{i}}
    \left(\sum_{i=0}^{m-1}\beta^{m-1-i}
    (\beta^{i}-\sum_{j=1}^{i}t_j\beta^{i-j}) \right)\\
     &=&\frac{1}{\sum_{i=0}^{m-1}\beta^{i}}
    \left( m \beta^{m-1}-\sum_{j=1}^{m-1}t_j
    \sum_{i=j}^{m-1}\beta^{m-1-j}\right)\\
     &=&\frac{1}{\sum_{i=0}^{m-1}\beta^{i}}
    \left( m
    \beta^{m-1}-\sum_{j=1}^{m-1}t_j(m-j)\beta^{m-1-j}\right)\\
    &=&\frac{p'(\beta)}
    {\sum_{i=0}^{m-1}\beta^{i}}=\frac{\beta-1}{\beta^m-1}p'(\beta).&
    \end{array}$$
    \end{proof}
    \begin{corollary}\label{ruzne_koreny}
    Let $\beta=\beta_1, \beta_2, \dots, \beta_m$ be mutually distinct roots of
    the Parry polynomial $p(x)$ of a~simple Parry number $\beta$. Then
     $$\lim_{n \to \infty}\frac{b_n}{n}=\frac{\beta-1}{\beta^m-1}
    \prod_{i=2}^{m}(\beta-\beta_i).$$
    \end{corollary}
    \begin{proof}
    $p(x)=\prod_{i=1}^{m}(x-\beta_i)$, $p'(x)=\sum_{k=1}^{m}\prod_{i=1, i \not =k}^{m}(x-\beta_i)$, thus
    $p'(\beta)=\prod_{i=2}^{m}(\beta-\beta_i)$.
    \end{proof}
    \begin{remark} If $p(x)$ is an irreducible polynomial, then $\beta$ is an algebraic
    integer of order $m$ and $\beta_2, \dots, \beta_m$ are algebraic conjugates of $\beta$,
    and hence mutually distinct.
    \end{remark}

\vspace{0.3cm} We now  study the asymptotic behavior of
$(b_n-c_{\beta}\, n)_{n \in \mathbb N}$. We know already that the
limit $\lim_{n \to \infty}\frac{b_n}{n}$ exists. Hence it is enough
to consider the limit of the subsequence $(U_n)$ $$\lim_{n \to
\infty}\frac{b_n}{n}=\lim_{n \to \infty}\frac{b_{U_n}}{U_n}=\lim_{n
\to \infty}\frac{\beta^{n}}{U_n}.$$ Under the assumption that all
roots of $p(x)$ are mutually different, we will find a~useful
expression for $U_n$. Since $M$ is diagonalizable, there exists an invertible matrix
$P$ such that  $PMP^{-1}$ is diagonal with
$\left(PMP^{-1}\right)_{ii}=\beta_i, \ i \in \{1, \dots, m\}.$
Using~\eqref{UiFormula}, we may write
\begin{equation}\label{U_i_new_formula} U_n=(1, \ 0, \dots, 0)P^{-1}\left(PMP^{-1}\right)^{n}P
\left(1 , 1, \dots, 1 \right)^T\, .
\end{equation}
It follows from the Perron-Frobenius theorem that $\beta>|\beta_i|$,
hence, the formula~(\ref{U_i_new_formula}) leads to the following
expression
\begin{equation}\label{c_new}
\frac{1}{c_{\beta}}=\lim_{n \to \infty}\frac{U_n}{\beta^{n}}=(1, \
0, \dots, 0)P^{-1}P_{e_1}P \left(1,1,\dots,1\right)^T,
\end{equation}
where $P_{e_1}$ is the orthogonal projection on the line given by
$e_1=(1,0,\dots,0)^T$, i.e., $\left(P_{e_1}\right)_{11}=1, \
\left(P_{e_1}\right)_{ij}=0 \ \text{otherwise}$. We now examine the
difference $b_n-c_{\beta}\, n$. Let $\langle b_n
\rangle_{\beta}=a_{k-1}\dots a_0\bullet $, thus
$b_n=\sum_{i=0}^{k-1}a_i \beta^{i}$ and $n=\sum_{i=0}^{k-1}a_i U_i.$
Employing~(\ref{U_i_new_formula}) and (\ref{c_new}), we obtain
$$\frac{1}{c_\beta}b_n-n=\sum_{i=0}^{k-1}a_i(1, 0, \dots, 0)P^{-1}\bigl(\beta_1^{i}P_{e_1}-
\left(PMP^{-1}\right)^{i}\bigr)P
\left(1, 1, \dots,1 \right)^T=$$
\begin{equation}\label{final_formula_simple}
\begin{array}{l}
    =(1, 0, \dots, 0)P^{-1}Z P \left(1, 1, \dots, 1 \right)^T,
    \end{array}\end{equation}
    where $Z$ is a~diagonal matrix with $Z_{11}=0, \ Z_{jj}=-z_j$ for $j \in \{2,\dots, m\}$,
    and $z_j=\sum_{i=0}^{k-1}a_i\beta_j^{i}.$  Since the coefficients
of $\beta$-expansion satisfy $a_i \in \{0,\dots, \lfloor \beta
\rfloor\}$ and since for PV numbers $\beta$, it holds $|\beta_j|<1$
for $j = 2,3,\dotsc,m$, we have
\begin{equation}\label{odhad_z_j_simple}
\vert z_j\vert \leq \sum_{i=0}^{k-1} \vert a_i \vert \vert
\beta_j^i\vert \leq \dfrac{\beta} {1 - \vert \beta_j \vert}.
\end{equation}
\begin{remark}\label{parry_minimal}
Suppose that the Parry polynomial $p(x)$ of a~Parry number $\beta$
is reducible, say $p(x)= q(x)\cdot r(x)$, where $q(x)$ is the
minimal polynomial of $\beta$, and $r(x)$ is a~polynomial of degree
at least~1. Then the product of the roots of $r(x)$ is an integer
and therefore either all roots of $r(x)$ lie on the unit circle or
at least one among the roots of $r(x)$ is in modulus larger than 1.
It implies that the set of $z_j$ is bounded for all $j$ if and only
if $\beta$ is a~Pisot number and its Parry polynomial is the minimal
polynomial of $\beta$.
\end{remark}
According to Remark~\ref{parry_minimal} and as $P$ does not depend
on $n$, we have shown the following theorem.
\begin{thm}
Let $\beta$ be a~simple Parry number. If $\beta$ is moreover a~Pisot
number and the Parry polynomial of $\beta$ is its minimal
polynomial, then $(b_n-c_{\beta}\, n)_{n \in \mathbb N}$ is
a~bounded sequence.
\end{thm}

Now $P$ is a~matrix of the Vandermonde's type given by
\begin{equation}\label{P_simple}
P_{ij}=\beta_i^{m-j},\ \text{then} \quad P \left(1, 1, \dots, 1
\right)^T=\left(\tfrac{\beta^m-1}{\beta-1},
\tfrac{\beta_2^{m}-1}{\beta_2-1}, \dots,
    \tfrac{\beta_m^{m}-1}{\beta_m-1}\right)^T.
\end{equation}
In order to have for all $n \in \mathbb N$ an explicit formula for
$\frac{1}{c_\beta}b_n-n$, it remains to determine $(1, \ 0, \dots,
0)P^{-1}$, i.e., the first row of $P^{-1}$. Since
$P^{-1}=\frac{1}{\det P}P^{adj},$ where $(P^{adj})_{1j}=\det P(j,
1)$ and $P(j,1)$ is obtained from $P$ by deleting the $j$-th row and
the $1$-st column,  applying Vandermonde's result yields
\begin{equation}\label{Pinverse_simple}(P^{-1})_{1j}=\frac{\prod_{i<k, \ i,k \not
=j}(\beta_i-\beta_k)}{\prod_{i<k}(\beta_i-\beta_k)}=\frac{(-1)^{j-1}}{\prod_{k\not
= j}(\beta_j-\beta_k)}=\frac{(-1)^{j-1}}{p'(\beta_j)}.
\end{equation}
Notice that since $p(x)$ does not have multiple roots,
$p'(\beta_j)\not=0$.

It follows from formulae~\eqref{final_formula_simple},
\eqref{P_simple}, and~\eqref{Pinverse_simple} that
$$\frac{1}{c_\beta}b_n-n=\sum_{j=2}^m \frac{(-1)^j z_j}{p'(\beta_j)}\frac{1-\beta_j^m}{1-\beta_j}.$$
In consequence, using the estimate~\eqref{odhad_z_j_simple}, we may
deduce an upper bound on $\vert b_n-c_{\beta}\, n \vert$
\begin{equation}
\label{upboundsp_simple} \vert b_n-c_{\beta}\,  n \vert \leq  \quad
2 c_\beta \beta \sum_{j=2}^m
\frac{1}{(1-|\beta_j|)^2}\frac{1}{|p'(\beta_j)|}\, .
\end{equation}

\begin{example}\label{Fib_example}
Let us illustrate the previous results on the case of  the simplest
simple Parry number - the golden mean $\beta=\tau = \frac{1 +
\sqrt{5}}{2}$. R\'enyi expansion of unity is $d_\tau(1)=11$ and
$p(x)=x^2-x-1$. The substitution matrix for the Fibonacci
substitution has been given in Example~\ref{Fib_matrix}.
Consequently, $(U_n)_{n \in \mathbb N}$ satisfies $U_n=f_n$ for all
$n \in \mathbb N$, where $(f_n)_{n \in \mathbb N}$ is the Fibonacci
sequence given by
$$f_{n+1}=f_n+f_{n-1}, \ f_0=1, \ f_1=2.$$

Applying Theorem~\ref{c_formula}, we get
$$c_{\tau}=\frac{\tau-1}{\tau^2-1}p'(\tau)=\frac{p'(\tau)}{\tau+1}=
\frac{2\tau-1}{\tau+1}=\frac{\tau^2+1}{\tau^3},$$ which is in
correspondence with Proposition~\ref{unit_gago}.

Let us denote the second root of $p(x)$ (the Galois conjugate of
$\tau$) by $\tau'$, $\tau' = \frac{1 - \sqrt{5}}{2} =
\tfrac{-1}{\tau}$. If $\langle b_n \rangle_\beta =a_{k-1}\dots
a_1a_0\bullet$, then $a_i \in \{0,1\}$ and
$$b_n-c_{\tau}\,n=c_{\tau}\,(\tfrac{1}{2\tau-1}, \tfrac{1}{2\tau'-1})\left(
\begin{array}{cc} 0 & 0 \\ 0 & -z_2\end{array}\right)\left( \begin{array}{c} 1+\tau \\ 1+\tau' \end{array}\right)
= \frac{1-\tau}{\tau(\tau+1)}\sum_{i=0}^{k-1}a_i (\tau')^{i}.$$
Since $|\tau'|<1$, the sequence $|b_n- c_{\tau}\,n|_{n \in \mathbb
N}$ is bounded and we may easily determine an upper bound (taking
into account that $\tau' <0$)
$$\frac{1-\tau}{\tau^3} \leq \frac{\tau-1}{\tau(\tau+1)}\sum_{i=1}^{\infty}(\tau')^{2i-1}
\leq \ c_{\tau}\,n-b_n \ \leq
\frac{\tau-1}{\tau(\tau+1)}\sum_{i=0}^{\infty}(\tau')^{2i}=\frac{1}{\tau^3},$$
thus, comparing the upper and lower bound, we deduce $$|b_n -
c_\tau\,n| \leq \frac{1}{\tau^3},$$ which is again in correspondence
with Proposition~\ref{unit_gago}, where we have replaced the
fractional part $\left\{ \tfrac{n+1}{ 1 + \beta}\right\}$ with 1 in
order to get an upper bound on $\vert b_n-c_{\tau}\,  n \vert$.
\end{example}
\subsection{Non-simple Parry numbers $\beta$}
Let $\beta$ be a~non-simple Parry number, then the R\'enyi expansion
of unity is of the form $d_\beta(1)=t_1t_2\dots t_m(t_{m+1}\dots
t_{m+p})^{\omega}$ with $m,p$ chosen to be minimal and $\beta$ is
the largest root of the Parry polynomial
$p(x)=(x^p-1)\left(x^{m}-t_1x^{m-1}-\dots
-t_{m}\right)-t_{m+1}x^{p-1}- \dots - t_{m+p-1}x-t_{m+p}$. Let us
recall that $p(x)$ may be reducible.

\vspace{0.3cm} Similarly to the simple case, our first goal is to
derive a~simple formula for constant $c_\beta$ such that $b_n
\underset{n \to \infty}{\approx} c_\beta n$. For any root $\gamma$
of the Parry polynomial $p(x)$,
$$(\underbrace{\gamma^{m-1}(\gamma^p-1), \ \gamma^{m-2}(\gamma^p-1),
\dots, (\gamma^p-1)}_{m \ \text{components}},
\underbrace{\gamma^{p-1}, \dots, \gamma, \ 1}_{p \
\text{components}})$$ is a~left eigenvector of the substitution
matrix $M$ associated with $\gamma$. On the other hand, according
to~(\ref{letter_frequencies}), the unique left eigenvector $(\rho_0,
\rho_1, \dots, \rho_{m+p-1})$ of $\beta$ such that
$\sum_{i=0}^{m+p-1} \rho_i =1$ satisfies that $\rho_i$ is the
frequency of letter $i$ in $u_\beta$. Combining the two previous
facts, we obtain for frequencies the following formula
\begin{equation}\label{frequency2}
    \rho_i=\frac{\sigma_i}{\sum_{i=0}^{m-1}\beta^{i}(\beta^p-1)+
    \sum_{i=0}^{p-1}\beta^{i}}=\frac{\sigma_i(\beta-1)}{\beta^m(\beta^p-1)},
    \end{equation}
where
$$\begin{array}{rclr}
\sigma_i & = & \beta^{m-1-i}(\beta^p-1) & \text{for }\ 0 \leq i \leq m-1, \\
\sigma_i & = & \beta^{m+p-1-i} & \text{for }\ m \leq i \leq m+p-1.
\end{array}$$

Let $(\Delta_0, \ \Delta_1, \dots, \Delta_{m+p-1})^T$ be the right
eigenvector of $M$ associated with $\beta$ such that $\Delta_0=1$.
Then it is easy to verify that $\Delta_i$ is the distance between
consecutive $\beta$-integers which is coded by letter $i$ in the
infinite word $u_\beta$ (see the formula for distances
in~\eqref{distance_beta}). Similarly as for simple Parry numbers,
also for non-simple Parry numbers the following formula for
distances holds and will be useful
\begin{equation}\label{distance2}
\Delta_i=\beta^{i}-\sum_{j=1}^{i}t_j \beta^{i-j}, \quad i \in \{0,
1, \dots, m+p-1\}.
\end{equation}
 \begin{thm}\label{c_formula2}
    Let $p(x)$ be the Parry polynomial of the non-simple Parry number $\beta$. Then
    \[c_{\beta}:=\lim_{n \to
    \infty}\frac{b_n}{n}=\frac{\beta-1}{\beta^m
    (\beta^p-1)}p'(\beta).\]
    \end{thm}
    \begin{proof}
    Similarly as for simple Parry numbers, $\lim_{n \to \infty}\dfrac{b_n}{n}$ exists and we have
    $$\lim_{n \to \infty}\frac{b_n}{n}=\rho_0\Delta_0+\rho_1\Delta_1+\dots+\rho_{m+p-1}\Delta_{m+p-1}.$$
    Applying~(\ref{frequency2}) and (\ref{distance2}), we obtain $\lim_{n \to \infty}\dfrac{b_n}{n}=\dfrac{\beta-1}{\beta^m
    (\beta^p-1)}\left(A+B\right)$, where
    $$A= \sum_{i=0}^{m-1}\beta^{m-1-i}(\beta^p-1)
    \bigr(\beta^{i}-\sum_{j=1}^{i}t_j\beta^{i-j}\bigr) \quad \text{and} \quad B=\sum_{i=m}^{m+p-1}\beta^{m+p-1-i}
    \bigl(\beta^{i}-\sum_{j=1}^{i}t_j\beta^{i-j}\bigr).$$
    It is then straightforward to prove that
    $$\begin{array}{c}A=(\beta^p-1)\bigl( m\beta^{m-1}-\sum_{j=1}^{m-1}
t_j(m-j)\beta^{m-j-1}\bigr),\\
\\
B=p\beta^{m+p-1}+p\sum_{j=1}^{m-1}t_j\beta^{m+p-1-j}+\sum_{j=1}^{p-1}(p-j)t_{m+j}\beta^{p-j-1}\\
\\
A+B=p'(\beta).\end{array}$$
    \end{proof}
    \begin{remark}
If we consider the infinite R\'enyi expansion of unity
$d_\beta^*(1)$ instead of the ``classical'' R\'enyi expansion of
unity $d_\beta(1)$, we have in the simple Parry case
$d_\beta^*(1)=\left(t_1\dots t_{m-1}(t_m-1)\right)^{\omega}$. Thus
the length $l$ of preperiod is 0 and the length $L$ of period is
$m$. In the non-simple Parry case, we have
$d_\beta^*(1)=d_\beta(1)=t_1t_2\dots t_m(t_{m+1}\dots
t_{m+p})^{\omega}$. Hence the length $l$ of preperiod is $m$ and the
length $L$ of period is $p$. With this notation, the formulae for
$c_{\beta}$ from Theorems~\ref{c_formula} and~\ref{c_formula2} may be
rewritten for both simple and non-simple Parry numbers in a~unique
way as
$$c_{\beta}=\frac{\beta-1}{\beta^l(\beta^L-1)}p'(\beta).$$
\end{remark}
    \begin{corollary}
    Let $\beta=\beta_1, \beta_2, \dots, \beta_{m+p}$ be mutually different roots of the
    Parry polynomial $p(x)$ of a~non-simple Parry number $\beta$. Then, $$c_{\beta}=\lim_{n \to \infty}\frac{b_n}{n}=
    \frac{\beta-1}{\beta^m
    (\beta^p-1)}\prod_{i=2}^{m+p}(\beta-\beta_i).$$
    \end{corollary}
    \begin{proof}
    Analogous as in Corollary~\ref{ruzne_koreny}.
    \end{proof}
    \begin{remark} If $p(x)$ is an irreducible polynomial, then $\beta$ is an algebraic integer of order
    $m+p$ and $\beta_2, \dots, \beta_{m+p}$ are algebraic conjugates of $\beta$, and hence mutually different.
    \end{remark}

\vspace{0.3cm} Let us now investigate the asymptotic behavior of $(b_n-c_{\beta}\,
n)_{n \in \mathbb N}$. As we know already that the limit $\lim_{n
\to \infty}\frac{b_n}{n}$ exists, we may rewrite it in terms of the
subsequence $(U_n)$
$$\lim_{n \to \infty}\frac{b_n}{n}=\lim_{n \to
\infty}\frac{b_{U_n}}{U_n}=\lim_{n \to
\infty}\frac{\beta^{n}}{U_n}.$$ Under the assumption that all roots
of $p(x)$ are mutually distinct, we will express $U_n$ in an easier
form. Since $M$ is diagonalizable, there exists an invertible  $P$ such that  $PMP^{-1}$ is diagonal with
$\left(PMP^{-1}\right)_{ii}=\beta_i, \ i \in \{1, \dots, m+p\}.$
Using~(\ref{UiFormula}), we may write
\begin{equation}\label{U_i_new_formula2} U_n=(1, \ 0, \dots, 0)P^{-1}
\left(PMP^{-1}\right)^{n}P \left(1,1,\dots,1\right)^T\, .
\end{equation}
It follows from the Perron-Frobenius theorem that $\beta>|\beta_i|$,
hence, the formula~(\ref{U_i_new_formula2}) leads to the following
expression
\begin{equation}\label{c_new2}
\frac{1}{c_{\beta}}=\lim_{n \to \infty}\frac{U_n}{\beta^{n}}=(1, \
0, \dots, 0)P^{-1} P_{e_1}P \left(1,1,\dots,1\right)^T\, .
\end{equation}
 Now, let us turn our attention to the
difference $b_n-c_{\beta} n$. Let $\langle b_n
\rangle_{\beta}=a_{k-1}\dots a_0\bullet $, thus
$b_n=\sum_{i=0}^{k-1}a_i \beta^{i}$ and $n=\sum_{i=0}^{k-1}a_i U_i.$
Employing~(\ref{U_i_new_formula2}) and (\ref{c_new2}), we obtain
$$\frac{1}{c_\beta}b_n-n=\sum_{i=0}^{k-1}a_i(1, 0, \dots, 0)P^{-1}\bigl(\beta_1^{i}P_{e_1}-
\left(PMP^{-1}\right)^{i}\bigr)P \left(1, 1, \dots,1 \right)^T=$$
\begin{equation}\label{final_formula_nonsimple}\begin{array}{l}
    =(1, 0, \dots, 0)P^{-1}Z P \left(1, 1, \dots, 1 \right)^T,
    \end{array}\end{equation}
    where $Z$ is a~diagonal matrix with $Z_{11}=0, \ Z_{jj}=-z_j$ for $j \in \{2,\dots, m+p\}$, and $z_j=\sum_{i=0}^{k-1}a_i\beta_j^{i}.$ Since the coefficients of
$\beta$-expansion satisfy $a_i \in \{0,\dots, \lfloor \beta
\rfloor\}$ and since for PV numbers $\beta$, it holds $|\beta_j|<1$
for $j = 2,3,\dotsc,m+p$, we have
\begin{equation}\label{odhad_z_j_nonsimple}
\vert z_j\vert \leq \sum_{i=0}^{k-1} \vert a_i \vert \vert
\beta_j^i\vert \leq \dfrac{\beta} {1 - \vert \beta_j \vert}.
\end{equation}
According to Remark~\ref{parry_minimal} and since $P$ does not
depend on $n$, we have shown the following theorem.
\begin{thm}
Let $\beta$ be a~non-simple Parry number. If $\beta$ is moreover
a~Pisot number and the Parry polynomial of $\beta$ is its minimal
polynomial, then $(b_n-c_{\beta}\,n)_{n \in \mathbb N}$ is a~bounded
sequence.
\end{thm}

The explicit form of the matrix $P$ reads
$$P=\left(\begin{array}{llllllll}\beta^{m-1}(\beta^p-1)&
\beta^{m-2}(\beta^p-1)& \dots & (\beta^p-1) & \beta^{p-1} & \dots &
\beta & 1\\
\beta_2^{m-1}(\beta_2^p-1)& \beta_2^{m-2}(\beta_2^p-1)& \dots &
(\beta_2^p-1) & \beta_2^{p-1} & \dots &
\beta_2 & 1\\
\vdots & \vdots & \ddots & \vdots & \vdots \\
\beta_{m+p}^{m-1}(\beta_{m+p}^p-1)&
\beta_{m+p}^{m-2}(\beta_{m+p}^p-1)& \dots & (\beta_{m+p}^p-1) &
\beta_{m+p}^{p-1} & \dots & \beta_{m+p} & 1
\end{array}\right)\, .$$
Hence,
\begin{equation}\label{P_nonsimple}
P \left(1,1,\dots, 1\right)^T=\left(\tfrac{\beta^m
    (\beta^p-1)}{\beta-1}, \ \tfrac{\beta_2^m
    (\beta_2^p-1)}{\beta_2-1},\ \dots, \ \tfrac{\beta_{m+p}^m
    (\beta_{m+p}^p-1)}{\beta_{m+p}-1}\right).
\end{equation}
In order to have for all $n \in \mathbb N$ an explicit formula for
$\frac{1}{c_{\beta}}b_n-n$, it remains to determine $(1, \ 0, \dots,
0)P^{-1}$, i.e., the first row of $P^{-1}$. By contrast to the
simple Parry case, the matrix $P$ is not in the Vandermonde's form.
However, we notice that its determinant is equal to a~Vandermonde
determinant through a~simple addition of columns. More precisely, we
start with the addition of the last column to the $m$-th column, the
last but one column to the $(m-1)$-st column and so forth. It is
readily seen that this procedure leads after $m$ steps to
a~Vandermonde matrix of order $m+p$ with the same determinant as
$P$.

So $\det{P} = \prod_{i<k}(\beta_i - \beta_k)$. The expression of
$(P^{-1})_{1j}, \ j \in \{1, \dots, m+p\}$, is then given by
\begin{equation}\label{Pinverse_nonsimple}(P^{-1})_{1j}=\frac{\prod_{i<k, \ i,k \not
=j}(\beta_i-\beta_k)}{\prod_{i<k}(\beta_i-\beta_k)}=\frac{(-1)^{j-1}}{\prod_{k\not
= j}(\beta_j-\beta_k)}=\frac{(-1)^{j-1}}{p'(\beta_j)}.
\end{equation}
Notice that since $p(x)$ does not have multiple roots,
$p'(\beta_j)\not=0$.

We obtain applying expressions~\eqref{final_formula_nonsimple},
\eqref{P_nonsimple}, and~\eqref{Pinverse_nonsimple}
$$\frac{1}{c_\beta}b_n-n=\sum_{j=2}^{m+p} \frac{(-1)^j z_j}{p'(\beta_j)}\frac{\beta_j^m(1-\beta_j^{p})}{1-\beta_j}.$$
In consequence, we may deduce an upper bound on $\vert
b_n-c_{\beta}\, n \vert$
\begin{equation}
\label{upboundsp_nonsimple} \vert b_n-c_{\beta}\,  n \vert \leq
\quad 2 c_\beta \beta \sum_{j=2}^m
\frac{1}{(1-|\beta_j|)^2}\frac{1}{|p'(\beta_j)|}\, .
\end{equation}

\begin{example}
Let us illustrate the previous results for the simplest non-simple
Parry number $\beta$ with R\'enyi expansion of unity
$d_\beta(1)=21^{\omega}$ and $p(x)=x^2-3x+1$, i.e., $\beta = \tau^2
= \tfrac{3 + \sqrt{5}}{2}$. The substitution matrix for the
associated substitution $\varphi: 0 \to 001, \ 1 \to
01$ is the square of the Fibonacci substitution matrix, i.e., $M_\varphi=\left(\begin{array}{cc} 2&1 \\
1&1
\end{array}\right)$.
Consequently, $(U_n)_{n \in \mathbb N}$ is just a~subsequence of the
Fibonacci sequence $(f_n)_{n \in \mathbb N}$ (defined in
Example~\ref{Fib_example}) given by $U_{n}=f_{2n}.$

Applying Theorem~\ref{c_formula2}, we get
$$c_{\beta}=\frac{\beta-1}{\beta(\beta-1)}p'(\beta)=\frac{p'(\beta)}{\beta}=
\frac{2\beta-3}{\beta}=1-\frac{1}{\beta^2},$$ which is in
correspondence with Proposition~\ref{unit_gago}.

Let us denote the second root of $p(x)$
by $\beta'$, $\beta' = \tfrac{3 - \sqrt{5}}{2}$. Let $\langle b_n \rangle_\beta =a_{k-1}\dots
a_1a_0\bullet$, then $a_i \in \{0,1,2\}$ and
$$b_n-c_{\beta}\, n=c_{\beta}\, (\tfrac{1}{2\beta-3}, \tfrac{-1}{2\beta'-3})\left(
\begin{array}{cc} 0 & 0 \\ 0 & -z_2\end{array}\right)\left( \begin{array}{c} \beta\\
\beta'
\end{array}\right)
= \frac{-1}{\beta^2}\sum_{i=0}^{k-1}a_i (\beta')^{i}.$$ Since
$0<\beta'<1$, the sequence $(b_n-c_{\beta}\, n)_{n \in \mathbb N}$
is bounded and we may easily determine an upper bound (taking into
account that coefficients in $\beta$-expansions satisfy the Parry
condition)
$$|b_n-c_{\beta}\, n| \leq \frac{1}{\beta^2}\left( 2+
\sum_{i=1}^{\infty}(\beta')^{i}\right)=\frac{1}{\beta^2}\left(2+\frac{\beta'}{1-\beta'}\right)=
\frac{1}{\beta^2}\left(2+\frac{1}{\beta-1}\right)= \frac{1}{\beta}\
,$$ which is in correspondence with the estimate we get if we
replace the fractional part $\left\{ \tfrac{n}{ \beta}\right\}$ with
1 in Proposition~\ref{unit_gago}.
\end{example}

\begin{description}
\item{\bf Open questions}:

It would be interesting to study the behaviour of $\beta$-integers
even for non-Parry numbers $\beta$, i.e., when the R\'enyi expansion
of unity of $\beta$ is not eventually periodic. Among non-Parry
numbers one may distinguish two cases:

$\bullet$ If the length of blocks of zero's in $d_\beta(1)$ is
bounded, say by a~length $L$, then the $\beta$-integers form
a~Delone set since the shortest distance between consecutive points
is at least $\frac{1}{\beta^L}$ and the largest distance is $1$.

$\bullet$ If $d_\beta(1)$ contains strings of zero's of unbounded
length, then the set of distances $(\Delta_k)$ between consecutive
$\beta$-integers have $0$ as its accumulation point, see
Equation~\eqref{distance_beta}. It means that in this case, the
$\beta$-integers do not form a~Delone set.

The first question to be answered in both cases is whether the limit
$b_n/n$ does exist for some of non-Parry numbers $\beta$.
\end{description}

\section*{Acknowledgements}

The authors acknowledge financial support by the grant MSM
6840770039 and LC06002 of the Ministry of Education, Youth, and
Sports of the Czech Republic.



{\small

\addcontentsline{toc}{section}{References}



\begin{thebibliography}{88}

\bibitem{jan} Janot, C.: Quasicrystals: a primer,
Oxford, Oxford University Press (1993)

\bibitem{IUC}
International Union of Crystallography, Acta Cryst. A 48 922-946
(1992)

\bibitem{BuFrGaKr}  Burd\' ik, \v C., Frougny, Ch., Gazeau, J. P.,
 Krejcar, R.: Beta-integers as natural counting systems for
quasicrystals, J. Phys A, Math. Gen. 31 6449-6472 (1998)


\bibitem{EFGV}
Elkharrat, A., Frougny, Ch., Gazeau, J. P., Verger-Gaugry, J. L.:
Symmetry groups for beta-lattices, Theor. Comp. Sci. 319 281-305
(2004)


\bibitem{FGK}
Frougny, Ch., Gazeau,  J.P., Krejcar, R.: Additive and
multiplicative properties of point-sets based on beta-integers,
Theor. Comp. Sci. 303 491-516 (2003)


\bibitem{GoGa} Gazeau, J.P., Verger-Gaugry, J. L.: Geometric study of the
beta-integers for a Perron number and mathematical quasicrystals, J.
Th. Nombres Bordeaux 16 1-25 (2004)

\bibitem{GaGo}
Gazeau, J. P., Verger-Gaugry, J. L.: Diffraction spectra of weighted
Delone sets on beta-lattices with beta a quadratic unitary Pisot
number, Annales de l'institut Fourier 56 2437-2461 (2006)

\bibitem{Re} R\'enyi, A.: Representations for real
numbers and their ergodic properties, Acta Math. Acad. Sci. Hungar.
8 477-493 (1957)

\bibitem{Pa} Parry, W.: On the beta-expansions of real numbers,
Acta Math. Acad. Sci. Hungar. 11 401-416 (1960)

\bibitem{Bertrand} Bertrand, A.: D{\'e}veloppements en base de Pisot et r{\'e}partition modulo {$1$},
C. R. Acad. Sci. Paris 285 419-421 (1977)

\bibitem{Th} Thurston, W. P.: Groups, tilings,
and finite state automata, Geometry supercomputer project research
report GCG1, University of Minnesota, USA (1989)

\bibitem{Qu} Queff\'elec, M.: Substitution Dynamical Systems -
Spectral Analysis, Lecture Notes in Math. 1294, Springer Berlin
(1987)

\bibitem{Fabre} Fabre, S.: Substitutions et {$\beta$}-syst{\`e}mes de
num{\'e}ration, Theoret. Comput. Sci. 137 219-236 (1995)

\bibitem{Lo1}
Lothaire, M.: Combinatorics on words, Encyclopedia of Mathematics,
Cambridge University Press (1983)

\end{thebibliography}}
\end{document}